# LOCATION- AND TIME-DEPENDENT VPD FOR PRIVACY-PRESERVING WIRELESS ACCESSES TO CLOUD SERVICES

Jong P. Yoon

MATH/CIS Department, Mercy College, Dobbs Ferry, New York, USA
jyoon@mercy.edu

## ABSTRACT

*The advent of smartphones in recent years has changed the wireless landscape. Smartphones have become a platform for online user interface to cloud databases. Cloud databases may provide a large set of user-private and sensitive data (i.e., objects), while smartphone users (i.e., subjects) provide location-sensitive information. Secure and private services in wireless accessing to cloud databases have been discussed actively for the past recent years. However, the previous techniques are unsatisfactory for dynamism of moving subjects' wireless accesses. In this paper, we propose a novel technique to dynamically generate virtual private databases (VPD) for each access by taking subjects' location and time information into account. The contribution of this paper includes a privacy-preserving access control mechanism for dynamism of wireless access.*

## 1. INTRODUCTION

Cloud computing models [1] give benefits to workflow management in various application domains where location varies and workflow covers vast wide areas. As a workflow is performed in a cloud, the location of field staffs (i.e., subjects) and items to deliver (i.e., objects) may change over the course of time, and the service quality may be determined by flexibility and automation for the location and time change. For example, global transportation and logistics industry carries out a workflow to deliver objects from an origin to the destination, perhaps with zero or more stop-over locations. One or more subjects, i.e., staffs in vessels or truck carriers, are also moving along the delivery route (see Figure 1). As such, a workflow is a sequence of tasks that can accomplish a business logistics process.

There is a group of subjects, i.e., service providers (SP), including the original sender of items, the final receiver and zero or more intermediate providers (or relayers) in between the sender SP and the receiver SP. There is another group of subjects such that service users (SU) request objects from a cloud. An SU can be an SP but at a different time with a different role. For example, when an object is relaying from one location to another, a subject as SP initially registers an object in a cloud database, and later the subject as SU may want to request the object [27,28].

The advent of smartphones in recent years has changed the wireless landscape. Smartphones, present in an ever increasing number of users, have become a platform for accessing cloud resources, where they are architected as client/server applications, and make use of the 4G connection to store data, and to provide/receive data to/from cloud databases. Vulnerability to compromise information becomes higher as smartphone accesses are increasing [21]. In enabling smartphones to access cloud resources, the ubiquity of social networks has led to compromising the data in cloud databases. In many situations, objects and subjects are moving at any point in time and location. Particularly, the moving subjects may wirelessly access a cloud database.

The services provided by SPs can be everything from the infrastructure, platform to software and data resources. Each such service is respectively called Software as a Service (SaaS), Platform as a Service (PaaS), or Infrastructure as a Service (IaaS). Depending on the service control granted to SU, the cloud services can be classified to one of the *cloud resources* (*) *as a service* (or *aaS). That is, in SaaS, SU has a limited admin control plus user-level controls only, while in PaaS, SU has the admin control and programming privileges for interfaces. In IaaS, however, SU has a total control on application, middleware, and guest operating systems, while PU still holds controls on hypervisor and hardware only. For example, Google Apps Engine (http://www.google.com/apps), Salesforce (http://www.salesforce.com), DropBox (http://www.dropbox.com) and Microsoft Dynamics are a **SaaS**, while Microsoft Windows Azure (http://www.microsoft.com/windowsazure) and Amazon's Beanstalk are a **PaaS**. Moreover, Amazon EC2 (http://aws.amazon.com/ec2), GoGrid (http://www.gogrid.com), Rackspace (http://www.rackspace.com), Rightscale (http://www.rightscale.com), and Joyent (http://joyent.com) are an **IaaS**.

Cloud services assume in this research that SPs provide SUs with digital resources about an object such as (cargo) items to be delivered and the resources are in {I|P|S}aaS. The shipment of delivery information systems consists of objects and the subjects, who have a control of the objects on a boat or a truck. <u>These subjects and objects are moving along the delivery route, and when wireless devices enable GPS, their location and time information is available to a cloud service</u> (see Figure 1). At the same time, there are also non-moving subjects who are either supporting or managing the shipment as shown in the upper layer of Figure 1. This figure illustrates a small subject of SPs and SUs who can play roles in cloud computing. There are numerous SPs such as suppliers, exporters, senders, relayers, custom clearance agents and freight forwarding agents. As illustrated in Figure 1, a resource (an object) is supplied by a manufacturer or an exporter (a subject), who may then grant a *privilege* to a new subject. From the beginning to the end of delivery, it is assumed that subjects and objects are available in a cloud, whereas they are moving from the origin location to the destination as shown in Figure 1. Along the transpositioning, there are carrier$_1$, carrier$_2$, …, carrier$_n$ (e.g., ships or trucks) and receiver in Figure 1. Note that subjects in the dotted red boxes are involved in a particular logistics and transportation process, and will be substantially discussed in Section 4 and 5.

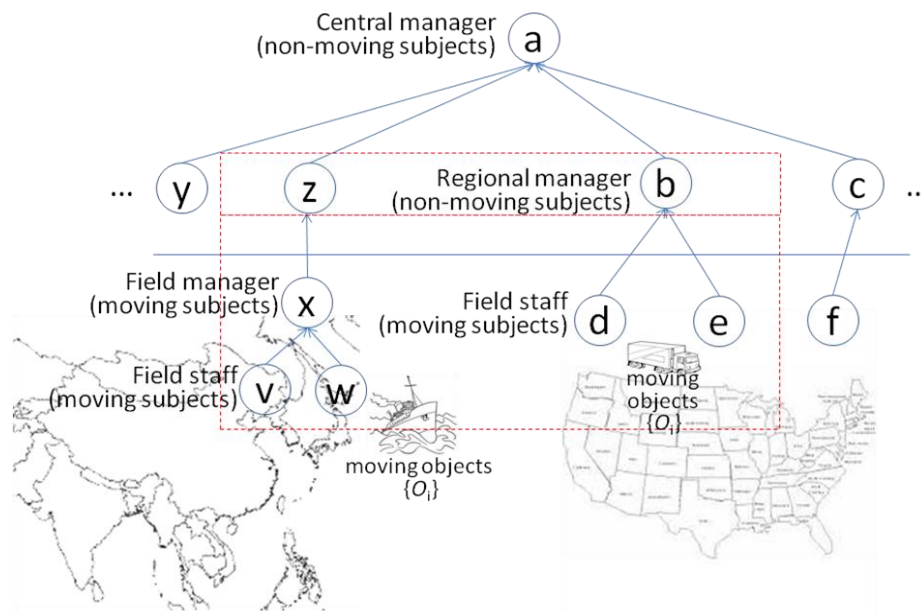

Figure 1: Moving Subjects and Objects
*Note that while moving, subjects may want to access objects in cloud databases, and their supervisors are changed based on location and time*

While objects are carried by carriers, the information about the objects change over the course of workflow tasks as shown in Figure 2. A set of variable-duration workflows is performed. Workflow "Importing" is a sequence of tasks, from "Registration" to "Door delivering", where some tasks have also subtasks, e.g., "Shipping-out" has "Port yarding" while "Ship assignment" followed by "Ship-loading." It is likely that over the course of workflow, location and time information of objects and subjects change. Such changes are not necessarily stored in a central cloud but sometimes stored in the smartphone memory or a local virtual machine. Therefore, the change of information is transmitted between smartphones and cloud servers, meaning that data is exchanged from one virtual machine (VM) to another. As such, an access privilege is granted to a SP/SU. Each subject has the different types of privilege on different granules of object. For example, a subject has an access privilege to update of the location of a moving object, while some other moving subject is granted to access objects of a cloud database. At some cases, a moving subject may want to access the cloud database about moving objects.

While the workflows are handled in a cloud database, there are important research issues which are not yet solved.

- A workflow is location-variant, meaning that a workflow performs different tasks based on location, so is the access privilege to objects. Objects are accessed by those moving subjects only if the location information stored in a cloud database is valid for access (e.g., a valid location is in Pacific Ocean). Typical access control mechanisms are based on pre-defined policies, and thus do not work efficiently in the situations of dynamism, where a subject is moving and its location is dynamically changed.

- A workflow is time-variant, and so is the access privilege to an object. The privilege permitted to a moving subject is effective only if the time information stored in a cloud database is valid for access (e.g., a valid time is in September 2010). Typical approaches, pre-definded policy-based access control mechanism, become inefficient due to this kind of dynamism in time.

- A workflow is organization-driven, meaning that a team of an organization is involved to carry out the (delivery) tasks (e.g., in Figure 1 and 2). Some of a team are moving, while the rest are not. Some of them have a smartphone-based access to a cloud server. The access privilege granted once should be revoked since the location and the time of subjects is out of the planned route and schedule. This dynamism of privilege grant and revocation based on subject's location and time has not considered successfully.

The goal of this project is to provide to each subject with a virtual private database (or *vpd* in short) which is dynamically generated depending on the location and time and so serve for moving subjects and objects. The contribution of this paper includes:

- The dynamism of access control by taking the request from moving subjects and objects is taken into consideration. The outcome of our work is a virtual private database (VPD). The characteristics of the VPDs for subjects are dynamically generated by their location and time, especially for smartphones access.

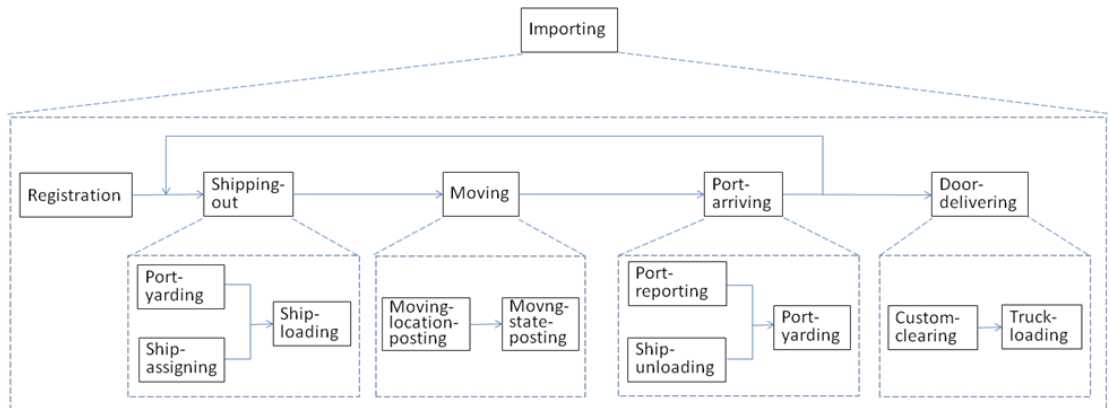
**Figure 2:** Workflow of "Importing"

- The VPD generated can preserve the privacy of subjects and both of their business management and social structures.

- Privileges granted and revoked are also reflected in the dynamic VPDs. Depending on location and time, a VPD is reconstructed automatically to grant or to revoke the privileges.

The remainder of this paper is organized as follows: Section 2 summaries the previous work. Section 3 describes authorization policies and proposes how to use the location and time information of smartphone accesses. Section 4 describes a method of generating VPD by appending the system context data which is loaded as a user accesses a cloud database. Section 5 describes about how VPDs are evolved or revoked as smartphone moves in or out from a task of workflows. Section 6 describes the conclusion of this paper.

## 2. RELATED WORK

One of the most challenging problems in managing operating system is the complexity of security administration. A naive and traditionally used access control to operating systems is the bi-factor authentication using login name and password [3]. This bi-factor cannot be used for authorization of cloud resources (or cloud databases).

A typical approach for authorization is access control mechanisms. Access control in cloud environments is provided using techniques such as VLANs and firewalls [2]. The techniques designed for enterprise environments and are ill suited for cloud computing environments, due to the high dynamism from multi-tenancy or possible attacks from internal tenancy. Access control is an indispensable component of operating system which mediates requests to resources of the system and makes decisions about whether or not they should be granted. Relative to Classical Discretionary Access control (DAC) [13], Mandatory Access Control (MAC) [14], Role-Based Access Control (RBAC) [15] model is more emphasized recently due to its simpleness, scalability, fine-grained control ability, and has been proven to be efficient to improve security administration with flexible authorization management.

Role based access control has become the predominant model for advanced access control since it reduces the cost of security management. There has been much work done to explore the role assignment, time constraint and security controlled mobility to enhance the network performance. In RBAC, users are assigned to roles, and permissions are granted to roles. The protection state is characterized by the triple <UA, PA, RR>, where UA is the user-role assignment relation, PA is the permission role assignment relation and RH is a role composition in systems. RBAC can greatly simplify the management of authorizations within a system, because a group of subjects are usually given the same permissions.

Odell and Parunak [15] found that an important characteristic of real-world systems is that the roles of subject may change over time. These changes can be of several different kinds. They analyze and classify the various kinds of role changes over time that may occur, and show how this analysis is useful in developing a more formal description of the application. These works provide guide for role transition among multiple domains in theory, however, they are not fit for dynamic role transition, especially for cloud computing.

Smartphones can also play a clone of the cloud computing [25]. In enabling smartphones to run in part cloud applications or to share any partitions of cloud resources, so called in an elastic computing, the vulnerability of such services will become higher. More security holes exist when not only static data but also control data is transmitted or exchanged between smartphones and cloud servers. For example, identity thieves can perform identity thefts more easily [26].

Some works consider role transition from temporal and spatial perspective [16,17,18], that is, roles of subject may change in different time periods and environments. Bertino et al. have proposed the Temporal-RBAC (TRBAC) model that addresses some of the temporal issues related to RBAC [19]. The main features of this model include periodic enabling of roles and temporal dependencies among roles which can be expressed through triggers. James et al. argues that TRBAC model addresses the role enabling constraints only. They propose a Generalized Temporal Role-Based Access Control (GTRBAC) model capable of expressing a wider range of temporal constraints [20]. In particular, the model allows expressing periodic as well as duration constraints on roles, user-role assignments, and role-permission assignments. Ray et al. [21] shows how RBAC can be extended to incorporate environmental contexts, such as time and location. The notion of RBAC has been revamped in the context of smartphone accesses [29]. The location information is added to the typical access control model. These works have exploited the users' identity as well as environmental parameters such as time and location, however, they depend on pre-defined authorization policies and therefore do not fit the dynamism of mobile accesses to cloud computing.

For remote access control, a few models have been proposed [22,23] which benefit from the advantages of both RBAC and trust management systems in an open environment. In particular, the TrustBAC model [22] supports automatic user-role assignment based on not only credentials of a stranger but its past behaviour and recommendations. Saffarian et al. propose a new dynamic user-role assignment approach for remote access control [24]. It addresses the principle of least privilege without degrading the efficiency of the access control system. What's more, it takes into account both credentials and the past behaviour of the requestor in such a way that he cannot compensate for the lack of necessary credentials by having a good past behaviour. However, this trustiness-based access control mechanisms do not fit successfully to the scenario such that a mobile user accesses a moving object (object to be delivered) in a cloud database.

Due to the uncertainty of execution time and task allocation, the methods mentioned above cannot fit access control well in operating system. We need to break the road from others: View-based access control mechanisms are developed based on hierarchically filtered views for file servers [6] and databases [7]. One of these approaches is fine-grained access control (FGAC), which provides row-level access granularity with a cost of query rewriting [8], or view creation [9]. Oracle VPD [10], an implementation of FGAC, defines policies as database functions attached to tables. Policies of this type require extra directives in the form of tables or views. Query rewriting is problematic in general [11].

To control mobile accesses to cloud databases, this proposed research utilizes the location and time information for both subjects and objects. This location and time information is automatically taken to use as a subject logs in a cloud system. A subject who is logged in from proper location at proper time can access proper objects. A set of the proper objects will be provided based on the relationship between the subject and the object.

## 3. AUTHORIZATION POLICIES

We propose a location and time-dependent access control mechanism. As subjects and objects are moving, their location and time data are used to determine authorization. Moving subjects may access wirelessly a cloud database, and thus the location and time information will be used to see if such information is valid. If valid, the access privilege is permitted to those moving subjects. Moreover, if an object is moving from one region to another, its supervision organization may be changed (e.g., supervision is moving from a regional manager to another).

In this section, the preliminaries are described. Authorization policy is defined. The policy can be extended by including location and time information of requesters (or subjects). As a requester may be moving from an origin to a destination, the location and time information for the requester can be verified and used to extend the authorization policy. Table 1 illustrates sample tables that will be used in the running examples.

### 3.1. Authorization Policy Model

In a policy-based access control mechanism, a policy manager creates and manages the policies that can be used to make access decisions. Typical authorization policies are defined over three elements, (subject, object, signed action), which means that subject is allowed to do action on

**Table 1**: Sample Tables in Cloud Database

subject

| id | name | Title | Specialty | Dept |
|---|---|---|---|---|
| s01 | Adam | Engineer | Electronic | IT |
| s02 | Alice | Manager | - | Trucking |
| s03 | Bob | Captain | - | Trucking |
| s04 | Parker | Driver | Furniture | Trucking |
| s05 | Charles | Manager | - | Operation |
| s06 | Chris | Manager | - | Delivery |
| … | … | … | … | … |
| s15 | Peter | Customer | Retailer | Wood |

assignment

| id | truck |
|---|---|
| s04 | t1 |
| s02 | t5 |
| s03 | t1 |
| s05 | t1 |

carrier

| id | Origin | Destination | Departure | Arrival |
|---|---|---|---|---|
| t1 | Vancouver | Miami | 08/11/2010 | 09/15/2010 |
| t5 | Anchorage | San Diego | 08/12/2010 | 08/21/2010 |

org_hierarchy

| OU | SubOU |
|---|---|
| Operation | Delivery |
| Delivery | Trucking |
| Operation | IT |

object

| oid | Name | sender | receiver | truck | Origin | Destination | Ship-out | Receive-in |
|---|---|---|---|---|---|---|---|---|
| o001 | Furniture | s11 | s21 | t1 | Seattle | Boston | 08/12/2010 | 08/31/2010 |
| o002 | Gold | s12 | s22 | t1 | Denver | New York | 08/19/2010 | 09/02/2010 |
| o003 | Car | s13 | s23 | t1 | Detroit | Miami | 08/22/2010 | 09/15/2010 |
| o004 | Metal | s14 | s24 | t1 | Pittsburgh | Washington | 08/29/2010 | 09/08/2010 |
| o005 | Lumber | s15 | s25 | t5 | Juno | Los Angeles | 08/12/2010 | 08/21/2010 |
| o007 | Rubber | s16 | S26 | - | Arlington | New York | - | - |

object. Depending on the sign of actions, subject is permitted to do the action if plus sign, or denied otherwise. The format of such policies as ($s$, $o$, $\pm a$), where $s$, $o$ and $\pm a$ respectively are denoted as subject, object and signed action [4,5].

The signed action specified for typical authorization policies is a privilege that can be applied to an object. This type of privileges is called *object privilege*. In addition to this, this paper proposes to use another type of privileges, which can be applied to a system. This type is called *system privilege*. Some examples of the system privileges are "grant" or "admin."

The *access control model* proposed in this research is also in ($s$, $o$, $\pm a$), but these three components, $s$, $o$ and $a$, are not pre-defined. The organizational hierarchy and workflow hierarchy dynamically change and their current states can obtained or derived from the cloud database. The detailed method will be described in the later section. The intuition behind our proposed method is that the current sailor of a vessel may be able to post the sailing states of the vessel to the database record of an object, only when the object is carried by that vessel at the current time. Using a location and timestamp of current users, a cloud database provides its proper subset of data to the eligible SPs and SUs.

There are two types of authorization policies: pre-defined and data-state-dependent policies, as illustrated in Figure 3. The pre-defined policies are defined with regardless of database states, while data-state-dependent policies are derived over data instances. One special case of pre-defined policies is the domain-independent policies. A pre-defined domain-independent policy governs as follows.

**Definition 3.1:** The authorization decision should satisfy all the domain-independent policies, logically noted as $\phi \models \psi$, where $\phi$ and $\psi$ denote the domain-independent policies and the rewritten SQL statements, respectively.

As wireless devices are popularly used to access cloud databases, the location and time information becomes one of the important factors to constitute the domain-dependent authorization policy. The data-state-dependent authorization policy will be described in the following section.

### 3.2. Location and Time to Extend Authorization Policies

The police "Subject Alice is permitted to read an object o005" is expressed in (Alice, o005, +r). Similarly, numerous policies are pre-defined as shown in Figure 3(a). Such policies are pre-defined for each case of user accesses to cloud databases. On the other hand, Figure 3(b) illustrates a data-state-dependent policy, meaning that the policy evolves as database states change. For example, an entity organization may change periodically.

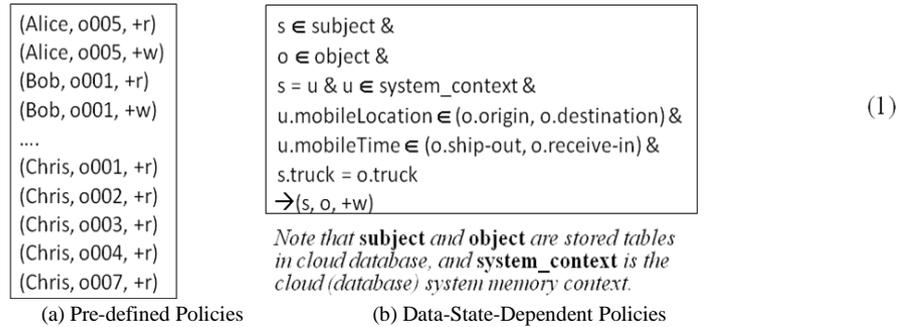

(a) Pre-defined Policies      (b) Data-State-Dependent Policies

**Figure 3:** Policy Examples

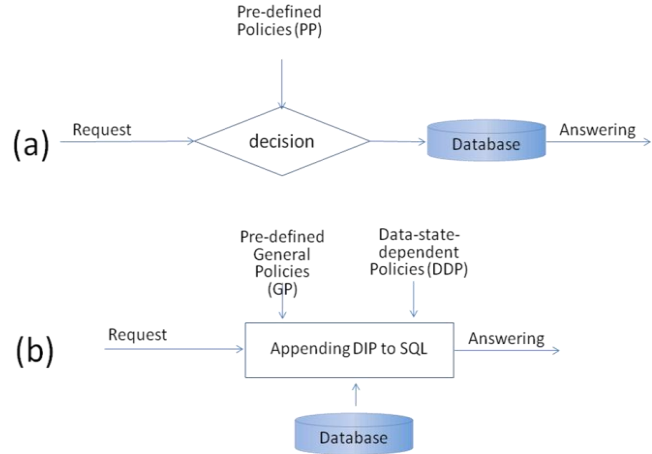

**Figure 4:** Access Control Mechanisms

If a subject accesses from a proper location (within the truck route), at a proper time (during the delivery time period), and also the subject is valid for access according to a pre-defined policy, the subject can be permitted to access a set of the objects. This data-state-dependent authorization policy states that a write access privilege is granted to a subject if the subject's session context is available in the system (perhaps in the virtual memory as illustrated in Figure 5), and if the subject's current location and time information is valid according to the cloud database.

Figure 4 illustrates how both pre-defined policies and data-state-dependent policies are enforced. In Figure 4(a), for a subject's request, pre-defined rules such as shown in Figure 3(a) are used to determine. For example, if Alice requests to access Object o005, then she is granted to read and write on the object. On the other hand in Figure 4(b), both general (pre-defined) policies and data-state-dependent policies are used. As an example, assume a general policy, (s, o, +w) → (s, o, +r), meaning that if a write privilege is granted, so is the read privilege. Together with this, finally domain-dependent policies are derived from the data-state-dependent policy. Suppose that Figure 3(b) is considered. While Parker is moving, if he requests to access an object, his location and time information is captured from his wireless device and loaded in the cloud (see system context in Figure 5). As his request is posed to a cloud database, his session information becomes available from the system context. Depending on the current database states being applied to Figure 3(b), it is derived that a write privilege is granted or denied. If granted, Parker can access to read the object.

## 4. CONSTRUCTING VIRTUAL PRIVATE DATABASES

The Virtual Private Database (VPD) is an example of fine-grained access control (FGAC) to modify user queries dynamically [8,9,10,11]. A VPD encodes the authorization policy into functions defined on each relation. Those functions, in conjunction with the user/application context, are used to generate predicates in the where clause to be appended to the user query before it is executed. The added predicates ensure that the user receives only those records in the table or view that are permitted by the authorization policy. In other word, VPD is an implementation of the query rewriting method for access control.

In this section, we use the query rewriting method not by appending the pre-defined authorization policy, e.g., Figure 3(a) [9], but by appending the current database state. The authorization policies are appended by general rule reasoning such as substitution and unification of rule variables and values. As an extension of access control mechanisms shown in Figure 4, our VPD approach is shown in Figure 6. When a moving subject requests an access, its location and time information together with the session user are loaded in a **system_context**

(e.g., GMM in Figure 5). The information available in the **system_context** is used to generate a VPD.

For example in Figure 3(b), subject Peter requests for an access. Since Peter, s15, is directly related with an object, Lumber, o005, the *vpd* for Peter is the object. With this intuition, we will conduct the following research activities.

The table **system_context** contains session parameters, e.g., session user, session login time, etc., and is available in the memory. In a cloud computing environment [12], there are a number of the memory spaces available, for example, smart phone memory, cloud server memory (or **gmm_context** in short) and the memory (or **vmm_context** in short) for a cloud database. As an example in Figure 5, the available main memory spaces are in a user wireless device, in a cloud server, and in a database system which is running on a guest platform. The **gmm_context** knows that the owner of a session is parker with his location and the login time. This session information lasts as long as the parker's session is on. The parker's request, say "select * from subject," will be rewritten as "select * from subject where id="parker"." It turns out that the only records of the table subject that are related with parker will be visible to the requester parker.

### 4.1. Functions to be Appended

Before applying the query rewriting method to generate a location-time-dependent VPD, we propose the following functions.

- *location_range*(), returns from the cloud database the range of the location information for the session user who is obtained in the **system_context**. For example, it returns the geocode (), e.g., latitude and longitude, of the origin and destination of the carrier on which a moving subject is.

- *time_range*(), returns from the cloud database the range of the time information for the session user who is obtained in the **system_context**. For example, it returns the time period of the departure and expected arrival time of the carrier on which a moving subject is.

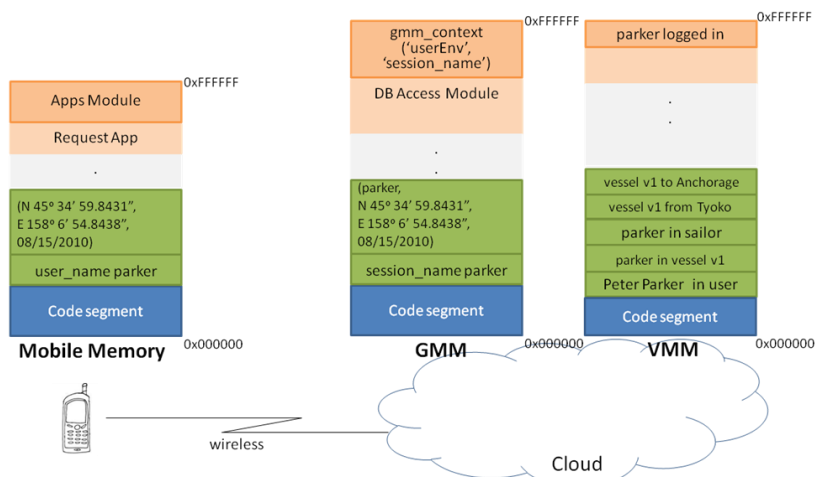

**Figure 5:** Session Context Example in Cloud Computing

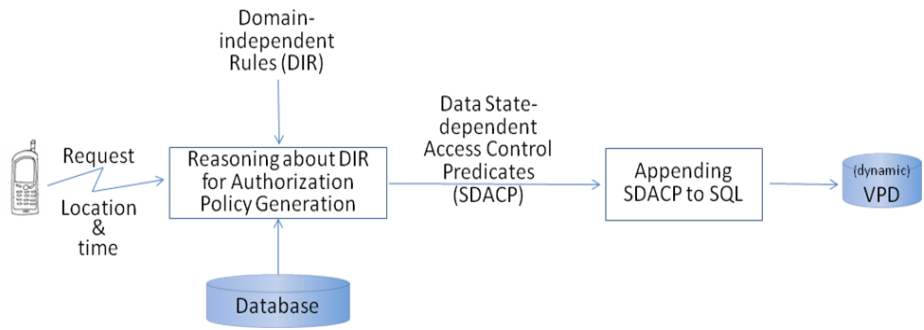

**Figure 6:** VPD for Wireless Access Control

- *workflow*(), takes a table (i.e., object table) to be accessed and returns the possible sequence of tables that can be linked from the subject table to the object table. For example, in Table 1, from the join conditions, `(subject.id = assignment.id) and (assignment.truck = object.truck)`, it is known that what subject is handling which objects.

- *organization*(), takes a pair of subjects and returns true if the first subject is subordinate to the second. For example, in Table 1, `(subject`$_1$`.dept = org_hierarchy.OU) and (org_hierarchy.sub_ou = subject`$_2$`.dept)` implies who works for whom who is along the organization hierarchy.

- *link*(), returns a conjunction of join predicates, which is formed from *workflow*() and *organization*().

One example of the possible ideas of using *organization*() is a policy that what can be accessed by subordinates can also be accessed by its manager or the manager of its superior. Similarly, *workflow*() specifies which carriers are a successor or a predecessor in a delivery route and henceforth who are on board as a wireless requester. Note that *location_range*() and *time_range*() are available from a cloud database. For example, in Table 1, the table truck contains such information. If a subject *a* is in truck t1, then his or her wireless access should appear along the route from Vancouver to Miami in the time period between August 11$^{th}$ and September 15$^{th}$, 2010.

### 4.2. Query Rewriting for VPD

The two functions, *range*() and *link*(), may be appended to a user request, say an SQL statement. Recall the functions *workflow*() and *organization*() and the main memory context called **system_context**. For given SQL statement, `select * from <table> where <condition>`, the rewritten SQL will be in the form

```
create vpd(subject) as
select  *
from    <table>
where   link() AND
        <condition>
```
(2)

where location and time are the information available in the system context as illustrated in Figure 5 (GMM or VMM). For the SQL, `select * from object`, requested by `Parker`, the rewritten SQL

```
create vpd(Parker) as
select * from subject, object
where  subject.name = sys_context:session_user AND
       subject.id = assignment.id AND
       assignment.truck = object.truck
```
(3)

returns objects {o001, o002, o003, o004}, since *link*() is appended. However, instead, *link*(subject.Specialty, object.Specialty) is appended, then the SQL is rewritten as

```
create vpd(Parker) as
select * from subject, object
where subject.Specialty = object.Specialty
```

returns {o001} due to the specialty of subject and the name of object are also used for joining operation.

Now, consider the location and time information, which is available **system_context**. Since in a logistics domain, the delivery route and also expected delivery time are planned, the function *range*() can return the range values efficiently well. Having the function, the following SQL has them appended.

```
create vpd(subject, location, time) as
select   *
from     <table>
where    location in range(<location>) AND
         time in range(<time>) AND
         link() AND
         <condition>
```
(4)

**EXAMPLE 4.1:** Suppose that `Parker` requests the SQL "`select * from object`". Since he sent the request wirelessly, the location and time information can be loaded in GMM as shown in Figure 4. Also, there is the table carrier is available as shown in Figure 5. With the *range*() function, the given query can be rewritten as

```
create vpd(Parker,l,t) as
select * from subject, object
where  sys_context:l in range(Parker, location)
       AND
       sys_context:t in range(Parker, time) AND
       subject.name = sys_context:session_user AND
       subject.id = assignment.id AND
       assignment.truck = object.truck
```
(5)

The above *vpn* is provided to `Parker` only if he is accessed in the valid location and time range. This is illustrated in Figure 7. In the figure, there are boxes, each has two elements, the upper small box indicates subject(s) *s* and the lower the *vpd*(*s*). If a box contains another, then the vpd of a container contains the vpd of its contained. For example, *vpd*(`Chris`) contains *vpd*(`Parker,l,t`) and *vpd*(`Bob,l,t`), meaning that Chris can access what Parker and Bob can.

Consider example tables of a cloud database in Table 1. Since `Charles` is the manager of `Operation` department, which is the superior organization to both `Delivery` and `IT` departments, in this example, according to the *link*() function, `Charles` can have an access to the objects to which `Chris`, `Alice` and `Adam` can, but not vice versa. In turn, `Chris` can access the *vpd* of `Parker`. The database `Parker` wirelessly access is *vpd*(`Parker,l,t`), which is shown in (5) above.

Recall Definition 3.1. As a VPD is constructed, the pre-defined domain-independent policies can be taken into account. See the following example.

**EXAMPLE 4.2:** Consider that `Chris` requests the same query as before: "`select * from object`". Assume that

$$\phi \supseteq \{\text{"The head of an organization can access the information of its subordinates"}\}.$$

(Note that this may be expressed in a simple form of logic, but left for readers.) According to the tables (Table 1), `Chris` is the manager of the `Delivery` Dept (in table subject), which contains `Trucking` Dept as its subordinate. Since his title is the head of the department, `Chris` can access all objects carried by both `Delivery` Dept and `Trucking` Dept. The subject set who work for `Trucking` Dept is {`s02`, `s03`, `s04`} and they are assigned to the vessels {`t1`, `t5`}. The objects carried by the vessels {`t1`, `t5`} are {`o001`, `o002`, `o003`, `o004`, `o005`}. It turns out that `Chris` can access the object set. This is written in SQL:

> **create vpd(Chris) as**
>
> ```
> select name
> from   subject s, assignment a, object o
> where  s.name = 'Chris' AND
>        s.id = a.id AND a.truck = o.truck
> ```
>
> **UNION** (6)
>
> ```
>  vpd(Parker,l,t)
> ```
>
> **UNION**
>
> ```
>  vpd(Bob,l,t)
> ```

which is equivalent to

> ```
> create vpd(Chris) as
>
> select name
> from   subject s, assignment a, object o
> where  s.name = 'Chris' AND
>        s.id = a.id AND a.truck = o.truck
> ```
>
> **UNION** (6')
>
> ```
> select name
> from   subject s, assignment a, object o
> where  s.Dept IN (select SubOU from org_hierarchy h
>                   where h.OU = s.Dept) AND
>        s.id = a.id AND a.truck = o.truck
> ```

We know that in the above rewritten SQL, $\psi$ satisfies $\phi$. Otherwise, it is not true that $\phi \models \psi$. Note that `Chris` is not a moving subject. However, since he is supervising Parker and Bob who are a moving object on the other hand as illustrated in Table 1 and Figure 7, the VPD granted to him is the above *vpd*(`Chris`). The privilege granted to `Chris` is to access the box labelled *vpd*(`Chris`) in figure, which include the *vpd*'s of `Parker` and `Bob`. Although *vpd*(`Chris`) is not direct location- and time-dependent, since its subordinate's VPDs, e.g., *vpd*(`Parker,l,t`) and *vpd*(`Bob,l,t`), are location-/time-dependent, Chris is indirectly affected as shown above (6) and (6'). .

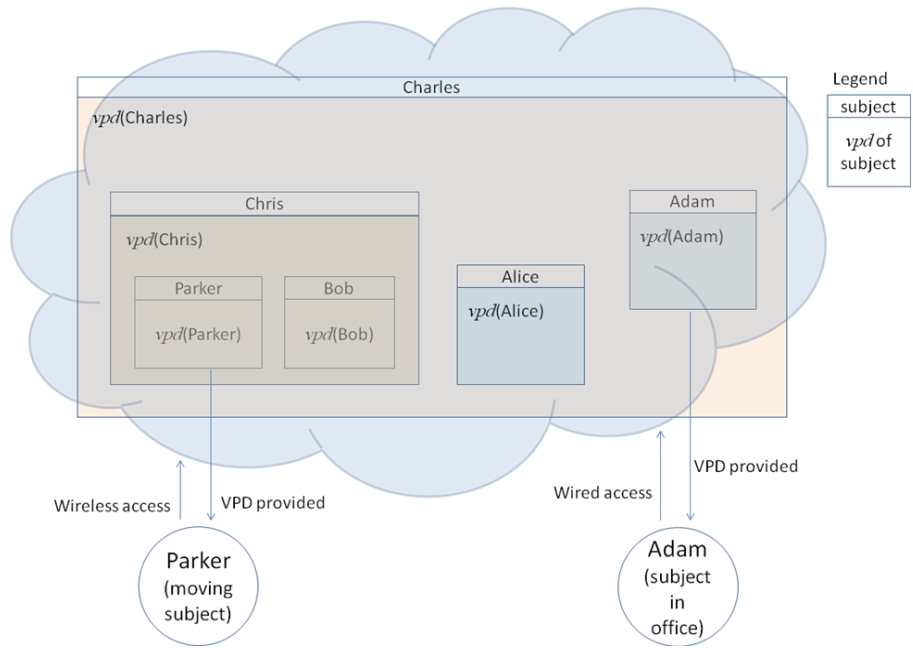

**Figure 7:** Wireless & Wired Access to VPDs

If the location and time information are not available or not in the range that computed by the server, the wireless request is denied or even the privilege once grated will be revoked. Depending on the general policy which may govern overall access control, only those objects that satisfy a subset of wireless subject's location and time range may be served. More issues about privilege granting and revoking are discussed in the following section.

## 5. GRANTING AND REVOKING PRIVILEGES

As a moving subject proceeds from an original site to a destination, the subjects and objects in a carrier are expected to be along the planned route. In general, when a subject enters in a delivery route, the subject is granted to an access privilege, and when exiting, the privilege is revoked. The privilege of accessing to a set of objects is formed in a virtual private database, i.e., *vpd*(*subject,location,time*). Therefore, *vpd*(*subject,location,time*) changes based on the location of a moving subject.

For granting and revoking privileges and therefore the generation of its VPDs, we extend the notion of VPDs discussed in Section 4. Location and time information about a subject is used to constitute the VPD for that subject.

**Definition 5.1:** Let the virtual private database granted to a subject *s* who is in the location *l* at the time *t* be *vpd*(*s,l,t*). *vpd*(*s,l,t*) is valid if there are in the cloud database an original location $l_o$ at the departure-time $t_b$ and a destination location $l_d$ at the arrival-time $t_e$ planned for *s*, such that $l_o \leq l \leq l_d$ and $t_b \leq t \leq t_e$. The privilege is granted to *s* as long as *vpd*(*s,l,t*) is valid. The privilege is revoked when *vpd*(*s,l,t*) is invalid.

For our example, recall Figure 1. There are subjects, some of which are moving, that is, `x, v, w, d, e, f` are moving objects. Assume that the subjects in the dotted red boxes are assigned to a workflow. The workflow delivers objects from Asia to USA, for example, where subjects, `x, v, w` are crew members of a ship from the origin, and these subjects are a subordinate of z. Figure 8(a) illustrates the VPDs such as *vpd*(z), *vpd*(x,*l*,*t*), *vpd*(v,*l*,*t*), and *vpd*(w,*l*,*t*). Since v and w are a moving subject in the same ship, *vpd*(v,*l*,*t*) ≈ *vpd*(w,*l*,*t*). If *vpd*(v,*l*,*t*) ∩ *vpd*(w,*l*,*t*) ≠ ∅, the privacy for v is in *vpd*(v,*l*,*t*) − *vpd*(w,*l*,*t*). Likewise, w's privacy is in *vpd*(w,*l*,*t*) − *vpd*(v,*l*,*t*).

Assume that the objects are handed over to trucks in Seattle, and the crew members x and w continue to move by the trucks. From there, since a new crew member d joined, all together three crews x, w and d are moving from Seattle at time $t_3$ to Chicago at $t_4$. The VPDs are illustrated in Figure 8(b). The moving subjects are x, w and d, where x has an access to *vpd*(w,*l*,*t*) but no access to *vpd*(d,*l*,*t*). It is true in part because they are subordinate to two different entities. Of course, since x and d are carrying the same objects, if they are indirectly related in the same global organization, the VPDs can be shared.

Anyhow, in this scenario, as a subordinate enters or exits in a workflow, its supervisor's VPD should be modified.

**Definition 5.2:** Let the virtual private database granted to a non-moving subject *s* who is in charge of its subordinates be *vpd*(*s*). *vpd*(*s*) is valid if there is no subordinate *r* such that *r* is a moving object and *vpd*(*r*,*l*,*t*) is invalid.

In Figure 8(b), the crew members x, w and d are supervised by z and b. The subject z and b can access the VPDs of their subordinate subjects x, w and d, but not vice versa. When x and w are out of the workflow, the valid VPD of b does not contain *vpd*(x,*t*,*l*) and *vpd*(w,*t*,*l*). Meanwhile, as the subject e enters in the workflow as a new crew member, the valid VPD of b contains *vpd(e,t,l)*. This is illustrated in Figure 8(c). In this way, the continuity of access [29] becomes possible in dynamic VPDs. In addition to that, this paper shows how organizational information can be used for the continuity of access designed in a VPD.

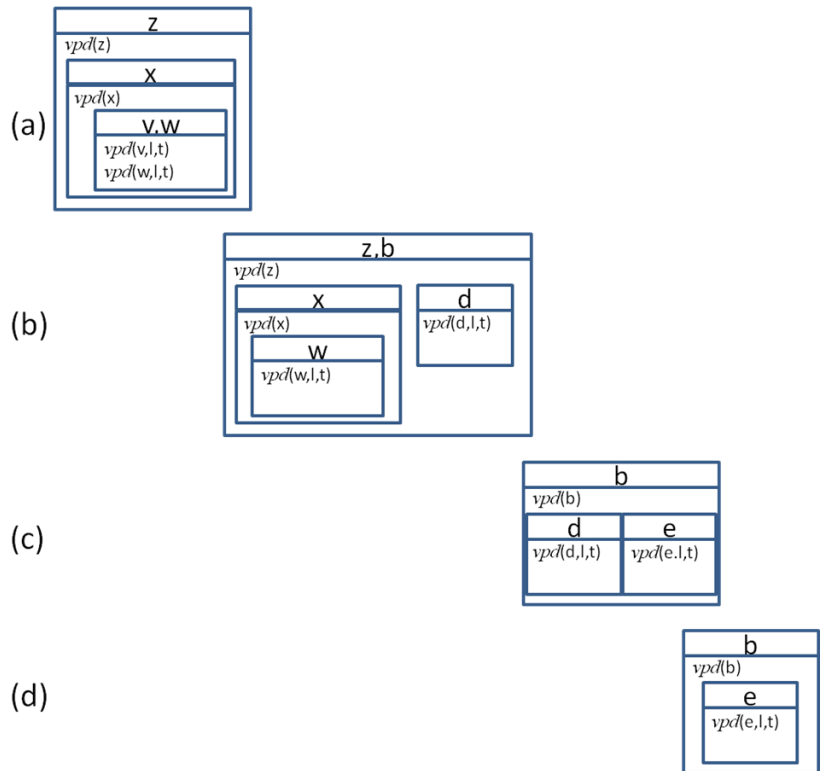

**Figure 8:** VPD Evolution for Moving Subjects over the Course of Workflow

Finally, subject `b` who is supervising `d` and `e` remains all the way to the destination. It is illustrated in the figure (d).

## 6. CONCLUSIONS

This paper describes the security and privacy management technique that can resolve the access control problems and difficulties arise when wireless accesses are allowed. The dynamism of wireless accesses is taken into consideration for setting up permitted database subsets. Such database subjects are automatically generated and granted to a wireless user as a virtual private database (VPD). Dynamic VPDs are derived by appending the location and time information from user accesses. Granting and revoking of the privileges for wireless accesses are controlled by dynamic VPDs. The techniques proposed in this paper are efficiently applied to the security and privacy management in workflow problem-domains where wireless devices are used to access cloud databases.

## ACKNOWLEDGEMENTS

The authors would like to thank everyone, just everyone!

**Authors**

Jong P. Yoon is a professor in Mercy College, MATH/Computer Science, New York. Dr. Yoon has published over 60 papers in various areas of Database Security, Information Security and Management, Web Technology. He has his Ph.D degree from George Mason University, Virginia, and MS from University of Florida.